# Kickstarting High-performance Energy-efficient Manycore Architectures with Epiphany

Andreas Olofsson†, Tomas Nordström*, and Zain Ul-Abdin*,

† Adapteva Inc.
Lexington, MA, USA
andreas@adapteva.com

*Centre for Research on Embedded Systems (CERES),
Halmstad University,
Halmstad, Sweden
{tomas.nordstrom, zain-ul-abdin}@hh.se

*Abstract*— In this paper we introduce Epiphany as a high-performance energy-efficient manycore architecture suitable for real-time embedded systems. This scalable architecture supports floating point operations in hardware and achieves 50 GFLOPS/W in 28 nm technology, making it suitable for high performance streaming applications like radio base stations and radar signal processing. Through an efficient 2D mesh Network-on-Chip and a distributed shared memory model, the architecture is scalable to thousands of cores on a single chip. An Epiphany-based open source computer named Parallella was launched in 2012 through Kickstarter crowd funding and has now shipped to thousands of customers around the world.

## I. INTRODUCTION

In recent years due to the never ending quest for more compute performance there an shift towards integration of many processor cores on a single chip, described as multicore or manycore architectures. Examples of commercial manycore efforts include the picoArray with 322 cores [1], Ambric with 336 cores [2], and Tile64 with 64 cores [3]. These processors were all designed for very specific markets, have power consumption that is an order of magnitude too high for most mobile embedded systems and do not have native support for floating point operations.

In this paper we set out to describe the efforts and results in designing a high-performance energy-efficient manycore architecture, supporting floating-point calculations, for use in embedded systems — *the Epiphany architecture*.

First we will describe the history and design goals for the Epiphany architecture. We also describe the creative financing used to kickstart a software eco-system around this new computer architecture. Then we describe the resulting architecture, its processor cores, memory and communication and some of the tradeoffs involved in the design process. We also describe the tools currently available and under development for Epiphany and take a look at some of the early applications developed on Epiphany. Finally, we draw some conclusions about future work anticipated for the Epiphany architecture and associated products.

## II. HISTORY

Andreas Olofsson worked at Analog Devices from 1998-2008, designing a number of application-specific processors including the TigerSHARC floating-point VLIW Digital Signal Processor (DSP) [4]. In 2004, the TigerSHARC was the most energy-efficient commercial floating-point processor in the market with 0.75 GFLOPS/W at 0.13um. Despite the impressive performance of the TigerSHARC, in actuality it was a very inefficient processor with only 2% of the silicon area used for floating-point math. A common industry assumption at the time was that inefficiencies in modern high level programmable processors are unavoidable and that the only solution is to use a radically different approach. Empowered by an intimate understanding of the real silicon cost of every feature in modern processors, Andreas founded Adapteva in 2008 with the mission to create an easy to program general purpose floating-point processor with an order of magnitude (>10X) better energy efficiency than legacy CPU solutions.

The following design goals were specified as "hard constraints" for the new architecture:

1. Energy efficiency (50 GFLOPS/W)
2. High raw performance (2 GFLOPS/core)
3. Scalable to thousands of cores
4. Easy to program in ANSI-C/C99
5. Implementable by a team of 5 engineers

In 2009, after a year and a half independent of research and development, Andreas announced the Epiphany manycore architecture during a panel presentation at the High Performance Embedded Computing Conference (HPEC) [5]. The Epiphany architecture was a clean slate design based on a bare-bones floating-point RISC instruction set architecture (ISA) and a packet based mesh Network-On-Chip (NOC) for effectively connecting together thousands of individual processors on a single chip. With a targeted power consumption of 1W per 50 GFLOPS of performance, this new architecture matched up well with the size, weight, and power (SWAP) constraints seen in modern state-of-the-art streaming signal processing system.

In May 2011, Adapteva introduced the first product based on the Epiphany architecture, a tiny 16-core 32 GFLOPS chip (E16G301) implemented in 65nm. Three months later in

August 2011, the team taped out a 64-core design (E64G401) in 28nm, demonstrating 50 GFLOPS/W (and 70 GFLOPS/W without IO).

Unfortunately, despite being the most energy-efficient floating-point processors in the world, the Epiphany chips were not gaining traction in any main stream high volume markets. Without a rich software and hardware eco-system built around the architecture, the transition cost was simply too high for most companies. All other parallel processor companies had run into similar challenges and had chosen to solve the problem by either independently building an eco-system through brute force spending or by providing complete application solutions for specific markets. Unfortunately, both of these approaches were extremely expensive, often amounting to tens of millions of USD in non-recoverable engineering expenses for the chip company.

Industry data has shown that it takes on average 8 years and costs between $100M-$400M for a public company and $25M-$100M for a startup to bring an advanced new processor solution to market. A rough breakdown of components contributing to the high development costs include (normalized to $100M): $10M for chip mask sets, $30M each for Applications software (software team) and Chip development (hardware team), and $10M for Software tool development (tools team), then $10M for EDA tool licensing, and finally $10M for Sales/Marketing.

When Adapteva launched in the midst of the 2008 recession, semiconductor startup venture funding had all but dried up, leaving Adapteva with very limited resources. As a result Adapteva was forced "to do more with less" and create a number of new design techniques aimed at cutting development costs. Cost-saving measures included iterative design simplification, a modular chip design methodology, horizontal engineering team integration, use of Multi-Project Wafers for product design, and the use of the open source tools such as Verilator and SystemC [6]. Using these cost-cutting techniques, Adapteva designed four generations of Epiphany manycore chips in less than 3 years with less than $2M in total investments, setting a new standard for cost-effective chip design in advanced technology nodes.

To tackle the challenge of building a complete parallel computing eco-system around Epiphany, Adapteva again innovated, turning to a crowd-funding platform called Kickstarter to finance development. In September 2012, Adapteva launched a $99 open source parallel computing project named Parallella with the dual goal of democratizing access to high performance parallel computing and building a software eco-system around the Epiphany architecture. The campaign was a huge success and in less than 30 days, Adapteva had raised close to 1M USD from 4,965 project backers, while committing to build and deliver over 6,300 Parallella computers to backers from 75 countries. All of the Parallella boards were eventually delivered in June 2014, making Adapteva the first semiconductor company in history to successfully crowd-fund development.

At the time of publishing, Adapteva has built over 20,000 Epiphany based Parallella computers and has delivered hardware to over 200 universities all around the world.

## III. ARCHITECTURE DESCRIPTION

The Epiphany architecture, shown in Fig. 1 and Fig. 2, consists of a 2D mesh of processor/mesh nodes ("eNode"), each including a 32-bit floating-point RISC CPU ("eCore"), multi-banked local memory, a direct memory access (DMA) engine, an event monitor, and a network interface. Each node connects to a NoC ("eMesh") through the network interface [7].

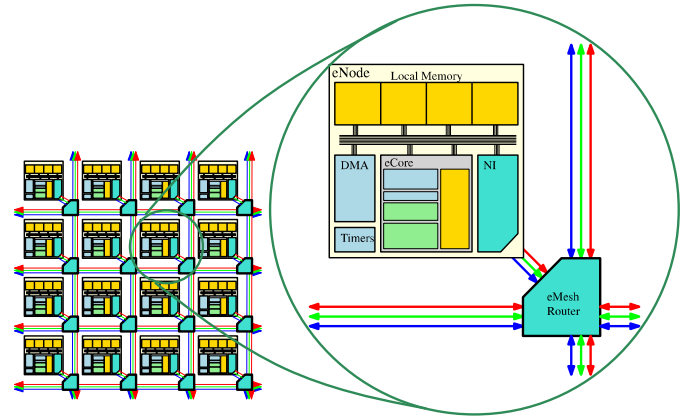

Fig. 1. The overall Epiphany architecture.

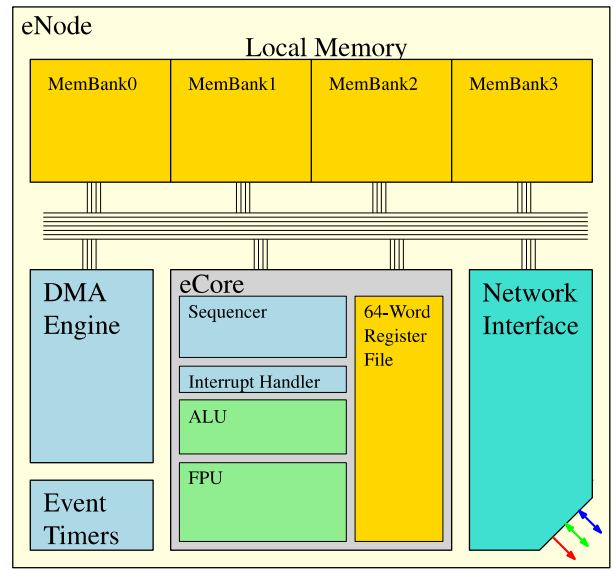

Fig. 2. The eNode Structure.

### A. Processing

The 32-bit Epiphany node processor "eCore" is an in-order dual-issue RISC architecture that includes an IEEE754 compatible floating-point unit (FPU), an integer arithmetic logic unit (ALU), and a 64-word register file.

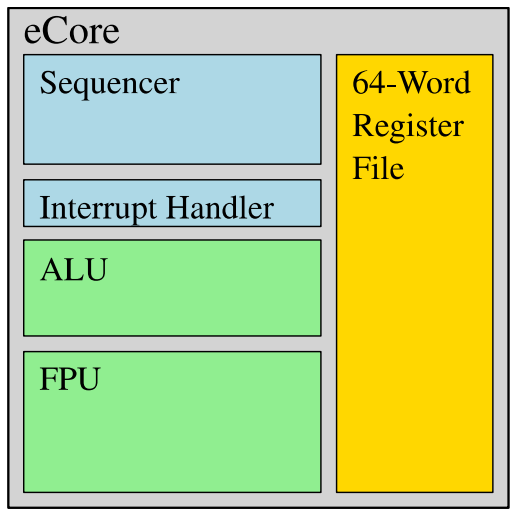

Fig. 3. The eCore Components.

*1) Instruction Set*

The Epiphany instruction set architecture (ISA) represents an ultra-compact RISC approach, with emphasis placed on real-time floating-point signal processing applications and C-programmability. The following list shows the complete set of instructions.

Epiphany Instruction Set:
- *Floating Point*: FADD, FSUB, FMUL, FMADD, FMSUB, FIX, FLOAT, FABS
- *Integer*: ADD, SUB, LSL, LSR, ASR, EOR, ORR, AND, BITR, NOP
- *Move*: MOV<COND>, MOVT, MOVFS, MOVTS, TESTSET
- *Load/Store*: (LDR, STR) * 3 addressing modes
- *Branching*: B<cond>, BL, JR, JALR
- *Core state and interrupt*: IDLE, TRAP, BKPT, RTI, GID, GIE, UNIMPL
- *Global (chip) state and synchronization*: SYNC, MBKPT, WAND

The architecture contains four flags set by integer operations (AN, AZ, AV, and AC) and two flags set by floating point operations (BN, BZ) [8]. These six flags and combinations of them result in 16 conditions codes that can be used for the conditional moves and branching. In addition, the floating-point operations set some sticky flags to indicate invalid (NAN) operands, underflow and overflow.

The Epiphany ISA implements variable 16/32-bit instruction encoding to maximize code density. All instructions are available as both 16-bit and 32-bit opcodes, with the instruction length being a function of the register number being used and the size of the immediate field. Instructions that use only R0-R7 and short immediate fields are encoded as 16-bit instructions.

*2) Register File*

The 9-port 64-word register file can in every cycle simultaneously perform the following operations:

- Three 32-bit floating-point operands can be read and one 32-bit result written by the FPU.
- Two 32-bit integer operands can be read and one 32-bit result written by the IALU.
- A 64-bit double-word can be written or read using a load/store instruction.

*3) Program Sequencer*

The eCore includes a program sequencer to support standard program flows including branches, jumps, calls, and function return, as well as nested interrupts. Each core contains a variable length 8-stage instruction pipeline. The first 5 stages (IF, IM, DE, RA, E1) are shared by all instructions and all but load/store and floating-point instructions are completed at the E1 stage. The data from the loads are written to the register file at stage 6 (E2), while the floating-point operations are completed at stage 7 (E3) or stage 8 (E4), depending on rounding mode. The instruction pipeline is fully interlocked to ensure the correct execution of sequential programs with data dependencies. The architecture can issue two instructions simultaneously provided that the two instructions to be executed do not have read-after-write or write-after-write register dependencies. Furthermore, only integer/load/store instructions are issued in parallel with FPU instructions.

*4) Interrupt Controller*

The interrupt controller provides full support for prioritized nested interrupt service routines. There are currently ten prioritized interrupts available: Sync/Reset (highest priority); Software exception; Memory fault; Timer0 expired; Timer1 expired; DMA0 complete; DMA1 complete; multicore wired and interrupt ("WAND"); and user interrupt (lowest). Every interrupt has a unique entry in the Interrupt Vector Table (IVT). The IVT is local to every core and is located at the beginning of local memory. All interrupts can be masked independently.

*B. Memory Architecture*

The Epiphany architecture does not include traditional L1/L2 hardware caching, opting instead to maximize local storage and memory bandwidth through the use of multi-banked scratchpad memory. Scratchpad SRAMs have much higher density than L1/L2 caches since they don't include complicated tag matching and comparison circuits.

The multi-banked local memory system supports simultaneous instruction fetching, data fetching, and multicore communication with four 8-byte-wide memory banks per core.

On every clock cycle, the following operations can occur:
- 64 bits of instructions can be fetched from memory to the program sequencer.
- 64 bits of data can be passed between the local memory and the CPU's register file.
- 64 bits can be written into the local memory from the network interface.
- 64 bits can be transferred from the local memory to the network using the local DMA.

The Epiphany memory architecture is based on a flat distributed memory map where each compute node is assigned a unique addressable slice of memory out of the total 32-bit address space. The overall memory is divided into blocks of up to 1MB that are co-located with each core in the multiprocessor system.

TABLE I. EPIPHANY ADDRESS SPACE.

| Bits | 31..26 | 25..20 | 19..0 |
|---|---|---|---|
| Address | Mesh Row | Mesh Column | Local |

The first 12 bits of an address specify a node by row and column, with the remaining 20 bits being local to that node, as shown in TABLE I. This means there can be at most 4096 mesh nodes operating within an addressable 64 x 64 2D mesh. Any memory region assigned to external DRAM must also be addressed through the same X/Y addressing scheme due to hard routing constraints imposed by the Epiphany NoC routing architecture.

A processor can access its own local memory and other processors' memory through regular load/store instructions. All read and write transactions from *local memory* follow a strong memory-order model. This means that the transactions complete in the same order in which they were dispatched by the program sequencer.

### C. Network-On-Chip

The communication in the Epiphany is supported by the eMesh Network-on-Chip (NoC), which consists of three independent 2D scalable mesh networks, each with four duplex links at every node. Each routing node consists of a round robin five direction arbiter and a single stage FIFO.

The routing mechanism is based on a distributed address-based routing that provides a single cycle wait, meaning that there is a single cycle routing latency per node. Each eMesh write link supports the transfer of 64 bits of data and 32 bits of address on each clock cycle.

Each of the three meshes have different purposes. Read requests travel through the rMesh, while cMesh and xMesh carry write transactions destined for on-chip and off-chip nodes, respectively. To the application, off-chip traffic is indistinguishable from on-chip traffic, apart from lower bandwidth and higher latency.

The eMesh architecture heavily favors writes over reads, since reading a foreign address involves sending a read request (over the rMesh) and waiting for the answer to arrive (on the cMesh). Writes, on the other hand, are of a fire-and-forget type, allowing the node to continue processing while the data moves through the NoC towards its destination.

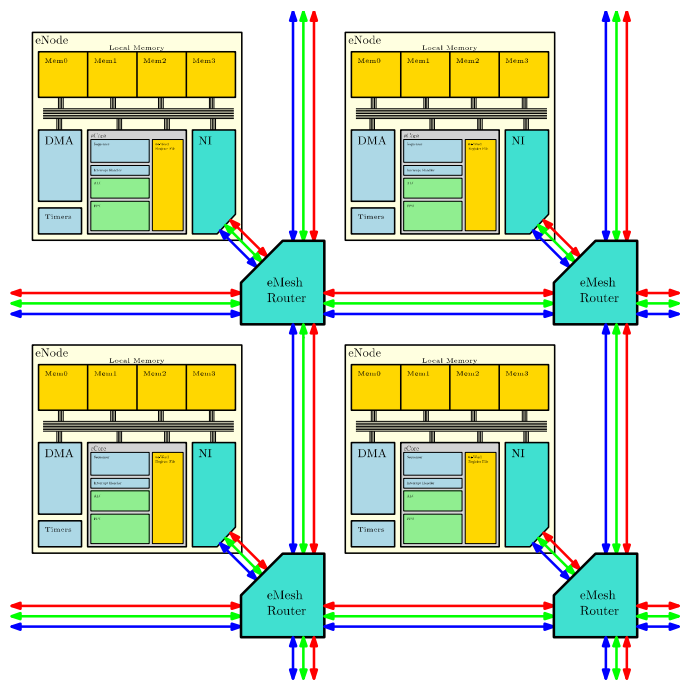

Fig. 4. The eMesh network consisting of three 2D meshes: read request (rMesh), on-chip write (cMesh), and off-chip write (xMesh).

The eMesh architecture heavily favors writes over reads, since reading a foreign address involves sending a read request (over the rMesh) and waiting for the answer to arrive (on the cMesh). Writes, on the other hand, are of a fire-and-forget type, allowing the node to continue processing while the data moves through the NoC towards its destination.

For read and write transactions that access *non-local memory*, the memory order restrictions are relaxed to improve performance. This relaxation of synchronization between memory-access instructions and their surrounding instructions is referred to as weak ordering of loads and stores. Weak ordering implies that the timing of the actual completion of the memory operations—even the order in which these events occur—may not align with how they appear in the sequence of the program source code. This non-determinism occurs for *read after write* to same or different locations, and *write after write to different locations*. If deterministic behavior is necessary within an application, ordering can be guaranteed through the use of software barriers.

Routing of traffic follows a few simple, static rules. At every hop, the router compares its own address with the destination address. If the row addresses are not equal, the packet gets immediately routed to the east or west; otherwise, if the column addresses are not equal, the packet gets routed to the north or south; otherwise the packet gets routed into the hub node, which then is the final destination. A read transaction consists of a read request on the rMesh and a write request on either the cMesh or xMesh.

When using the multicast feature of the mesh, a different routing algorithm is used instead. In this case, the data is sent radially outwards from the transmitting node. All nodes compare the destination address with a local multicast register, and if both values are equal (the node is listening to that traffic), it enters the node. This feature allows writing to multiple nodes using a single transaction.

Key features of the on-chip mesh network include:
- *Optimization of write transactions over read transactions*. Writes are approximately 16x more efficient than reads for on-chip transactions. Programs should use the high write-transaction bandwidth and minimize inter-node, on-chip read transactions.
- *Separation of on-chip and off-chip traffic*. The separation of the xMesh and cMesh networks decouples off-chip and on-chip communication, making it possible to write on-chip applications that have deterministic execution times regardless of the types of applications running on neighboring nodes.
- *Deadlock-free operation*. The physically independent read and write meshes—together with a fixed routing scheme of moving transactions first along rows, then along columns—guarantees that the network is free of deadlocks for all traffic conditions.
- *Scalability*. The implementation of the eMesh network allows it to scale to very large arrays. The only limitation is the size of the address space. A 32-bit Epiphany architecture scales to 4,096 processors and a 64-bit architecture scales to 18 billion processing elements within a unified shared memory system.

*D. External I/O*

The on-chip eMesh network extends off-chip through four (North, East, West, South) multiplexed (byte-oriented) I/O links. Each off-chip link provides bidirectional bandwidth of 1 GB/s. These chip-to-chip links fully support the internal push-back scheme and are designed to efficiently interface with low cost FPGAs. The links can also be used to construct larger glue-less arrays of up to 4,096 cores on a single board using a large number of Epiphany based chips arranged in 2D grid.

## IV. DESIGN DECISIONS

The following section provides insight into some of the design decision made during the development of the Epiphany architecture.

*A. CPU Architecture Tradeoffs*

The eCore CPU design philosophy was to be "as simple as possible but not simpler." In the domain of massively parallel computers there have been a number of architectures with impressive performance numbers that were very difficult to program. The design goal of the Epiphany was to offer a 10x energy efficiency boost over legacy CPU architectures while still supporting an ANSI-C programming flow that the average programmer could easily master.

To realize the firm design goals of a high-performance, easy to use ANSI-C/C99 programmable machine, the following C-friendly features were included in the Epiphany architecture:
- In-order dual issue scheduling to boost performance for virtually all applications.
- Native IEEE floating-point instructions to let programmers focus on the mathematical algorithms instead of operand scaling and numerical precision.
- A large 64-entry register file to enable efficient loop unrolling and efficient variable reuse.
- A 64-bit load/store path to boost performance for many memory-bound math algorithms.
- Native byte addressability to support true ANSI-C/C99.
- Low cost features like the interrupt controller, debug unit, and timers to make the manycore Epiphany processor feel more like a traditional CPU.

Any features that clashed with the original five design goals set out for the Epiphany architecture were dismissed. Examples of rejected features included:
- Out of order scheduling would not have given a significant performance boost in streaming signal processing applications and clashed with the energy efficiency design goal.
- Hardware caching would have made reaching energy efficiency target of 50 GFLOPS/W impossible and it also would have made scaling to thousands of cores very difficult due to the complexity of coherency and off-chip DRAM access contention.
- Bit manipulation instructions such as Not, Mask, Rotate, Add Carry, and Ones were left out because they are rare in signal processing and would have incurred an incremental but unacceptable power and area penalty in every active clock cycle.
- Orthogonal data type ISA support was left out in order meet the energy efficiency target. One of the key takeaways from the TigerSHARC was that the cost of an orthogonal ISA is enormous in terms of hardware design complexity. It is far more efficient to have a base architecture with a small set of derivative instruction sets for different markets (i.e. signed, unsigned, 8b, 16b, 32b) data type ISA.
- A branch target buffer was not implemented, as this would likely have made reaching the energy efficiency goal impossible. Instead a simple scheme of "branch never taken" was chosen with a 3 cycle branch penalty imposed for every branch that is taken.

*B. Network-on-chip Tradeoffs*

The design of the Epiphany network-on-chip also involved a number of non-obvious design tradeoffs. For Epiphany the reasoning around these tradeoffs can be structured around selection of topology, packet- or circuit-switched network, and finally routing and flow control.

*1) eMesh topology*

For Epiphany, a 2D mesh network was selected as this is a simple topology that matches up well with the planar layout topology used for standard CMOS processes and is well understood after decades of research. Furthermore, many of the intended math and signal processing algorithms have already been mapped to 2D mesh networks. Torus wraparound connections were ruled out as they would have required twice as many wires per cross-section. Higher order networks such as butterfly and CLOS were ruled out as being too large and too complex to be suitable considering the traffic patterns found in the signal processing applications.

*2) Packet switching*

Address-based packet switching scheme was chosen in place of a circuit switch-based network to provide flexibility across a broad set of applications. The most important and non-intuitive design decision for the eMesh NoC has been to send a complete destination address with every data transaction. The hardware saved through the resulting design simplification far outweighed the cost of the extra bits of address transmission. The design choices may seem counter-intuitive to those familiar with large scale networks, because the relative cost of wires, drivers, and registers are completely different for NoCs compared to larger system-based networks where wire costs tend to be extremely high.

*3) Routing and flow control*

The eMesh routing is based on a simple state-less X/Y routing enabling scaling to large array sizes. Flow control is handled through a pushback signal propagating backwards through the mesh at the same speed that the traffic moves forward, ensuring that no packets are ever lost. The single cycle push back flow control method cost an extra stage of shadow registers and a mux at every routing stage, but this cost was deemed acceptable compared with the benefit of single cycle hop latencies and low stall penalties.

QOS features were kept out of the first generations of the Epiphany architecture because it was assumed that most traffic patterns in signal processing are known a-priori and well-behaved. Due to strict real time requirements found in certain applications, the issue of QOS is now being revisited.

## V. EPIPHANY CHIP IMPLEMENTATION

The Epiphany architecture has been implanted in four separate chips, culminating in the latest 4$^{th}$ generation 28 nm Epiphany-IV 64-core design. The Epiphany-IV contains 64 identical floating point eCores, each one with 32 KB of local memory.

The Epiphany-IV was implemented in a 28 nm LP CMOS process with 9 layers of copper metal for routing and contains over 160 million transistors in an area of 10 mm$^2$. The chip was implemented in hierarchical tiled layout methodology, wherein a basic processor tile was implemented as a complete hard macro with all pins and clocks completely exposed at the four sides of the tile. The top level chip layout was done completely through abutment and without any signal routing. Any signal that needed to be communicated across the chip, was registered within each tile and fed through to its neighbors. The Epiphany-IV was implemented in less than 12 weeks by 2 engineers, translating to an efficiency of 1 M transistors per engineer per day.

The energy efficiency of the Epiphany chips are mostly a byproduct of architectural design decisions rather than circuit level optimization. No custom logic was used in implementation and off the shelf 9-track standard cell libraries were used to implement the design. Leakage power was minimized by not using LVT threshold transistors while dynamic idle power was minimized through extensive use of clock gating and power saving idle modes.

The Epiphany-IV has been operated up to 800 MHz in the lab, offering a peak theoretical performance of 102 GFLOPS (while drawing 1.74 W). It theoretically has 102 GB/s of bisection bandwidth, 1.6 TB/s of on-chip local memory bandwidth, and 7.2 GB/s of aggregate off-chip bandwidth. The peak energy efficiency of of 70 GFLOPS/W is achieved at 0.9 V and a frequency of 500 MHz.

## VI. THE PARALLELLA BOARD

The most prominent usage of the third generation Epiphany silicon E16G301 is the Parallella board [9]. The development of this board was financed in 2012 as a Kickstarter project.

The Parallella board, shown in in Fig. 5., is a fully open-source credit-card sized computer containing a 16-core Epiphany E16G301, a Xilinx Zynq 7010/7020, and 1 GiB of RAM. The Xilinx Zynq is a System-on-Chip (SoC) with two ARM Cortex-A9 processor cores and some reconfigurable FPGA logic and is fully supported by Linux. The board also contains GBit-Ethernet, USB and HDMI interfaces, can boot from a MicroSD card, and is able to run the Epiphany SDK.

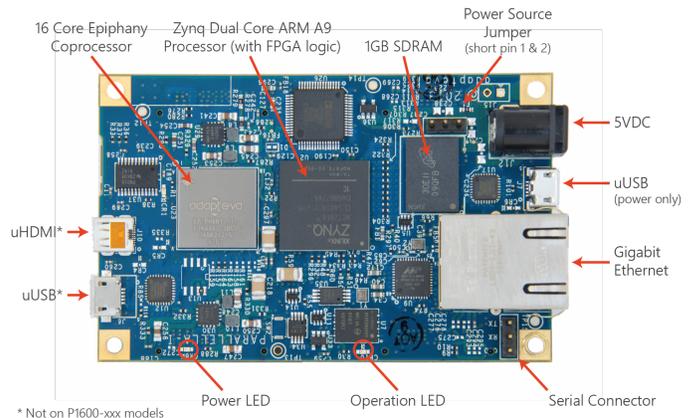

Fig. 5. The Parallella board.

Adapteva's e-Link interface is implemented inside the FPGA logic and is used to exchange data between the ARM cores and the Epiphany. By default, a 32 MiB block of memory is shared between both systems and starts at address 0x8e000000, which translates to mesh coordinates between (35,32) and (36,0). The 4x4 grid of processor nodes uses the coordinates between (32,8) and (35,11) inclusive.

To allow for daughter boards, four expansion connectors are provided, allowing access to the power supplies, general I2C, UART, GPIO and JTAG interfaces as well as the northern and southern eLink interfaces to the Epiphany chip.

## VII. PROGRAM DEVELOPMENT

For the Epiphany architecture Adapteva has released a fully open source software development kit (SDK) [10]. It provides a standard C programming environment based on the GNU toolchain (gcc, binutils, gdb) and the newlib [11] and allows the execution of regular ANSI-C programs on the Epiphany cores. Each core in the architecture runs a separate program, which is built independently and then loaded onto the chip by a host processor using a common loader. An Eclipse-based integrated design environment manages the details of configuring and building the project for the manycore architecture.

Architecture specific features like DMAs and the interrupt controller can be programmed through he Epiphany Utility Library (eLib) API. The eLib library also contains support for parallel computing synchronization operations like mutexes and barrier functions.

For the host processor an Epiphany Host Library (eHAL) provides direct access to the eCores, for loading programs, starting, resetting, and message passing. It furthermore provides read and write access to both the cores and the shared DRAM.

On the host, all normal Linux operating system functions are available, and any programming language able to interface to the available C libraries may be used to control the Epiphany. The eCores themselves don't run any operating system, but provide a "bare metal" standard C environment for programming.

Thanks to the recent Parallella project, the Epiphany architecture can now be programmed with a number of different APIs and frameworks. Brown Deer Technology has extended its open source GPL licensed COPRTHR library to add Epiphany support for OpenCL, STandarD Compute Layer (STDCL), and bare metal coprocessor threads. The COPRTHR SDK targets heterogeneous platforms such as CPUs, GPUs, and Parallella [12]. The support for Erlang functional programming language is provided by developing an application framework that allows executing OpenCL kernels within nodes and binding it with Erlang network interface modules [13]. The Array manipulation language (APL) is compiled by the APL to C compiler [14] being developed at Lab-Tools Ltd. We have implemented a backend code generator in our in-house CAL tool-chain for CAL dataflow language compilation targeting the Epiphany architecture [15]. Additional open source efforts are currently underway to create OpenMP and MPI programming frameworks for the Parallella.

## VIII. EXAMPLE APPLICATIONS

We have evaluated the 16-core Epiphany processor by implementing two significant case studies, namely; an autofocus criterion calculation and the fast factorized back-projection algorithm, both key components in modern synthetic aperture radar systems [16]. One of the Epiphany implementations demonstrates the usefulness of the architecture for the streaming based algorithm (the autofocus criterion calculation) by achieving a speedup of 8.9x over a sequential implementation on a state-of-the-art general-purpose processor of a later silicon technology generation and operating at a 2.7x higher clock speed.

We have also evaluated our approach of compiling CAL to parallel C code with support for message passing being implemented as a stand-alone library by using 1D-DCT algorithm [17]. Our preliminary results reveal that the hand-written implementation has only 1.3x better throughput performance with respect to the auto-generated implementation, which is quite competitive. Also, the CAL high-level language approach leads to reduced development effort.

Another significant study done on the Epiphany architecture is the implementation of the bcrypt cryptography algorithm [18]. The bcrypt implementation for Epiphany is optimized in assembly and executes two bcrypt instances per core. The 64-core Epiphany implementation achieves 2400 cycles per second per Watt (c/s/W) compared to 79 c/s/W for the Intel i7-4770K CPU and 46 c/s/W for the AMD HD7970 GPU. These energy efficiency results (measured as cycles per second per watt) reveal that Epiphany outperforms the Intel CPU and AMD GPU by a factor of 30x and 52x respectively.

A five-point star-shaped stencil-based application kernel was implemented using C and assembly on a 64-core 600 MHz Epiphany-IV development platform, reaching a sustained performance of 65 GFLOPS (85% of peak) [19]. The implementation was able to achieve on-chip DMA bandwidth of 2 GB/s, off-chip memory access bandwidth of 150 MB/s, and a normalized power efficiency of 32 GFLOPS/W.

## IX. FUTURE EPIPHANY WORK

The Epiphany architecture was designed for embedded real time signal processing applications, but due to its impressive energy efficiency and scalability, the Epiphany has also generated significant interest within the field of High Performance Computing (HPC). To address the HPC market, future Epiphany products will need HPC-centric extensions for 64-bit floating-point operations, 64-bit addressing, memory fault tolerance (ECC), and 128 KiB of memory.

To reduce I/O bottlenecks, the existing 1 Gbps/lane source synchronous LVDS eLink interfaces will be replaced with multiple high speed serial 10 Gbps/lane interfaces.

## X. CONCLUSIONS

To summarize, we have introduced Epiphany, a processor suitable for high-performance embedded systems. General purpose programmability and the outstanding energy efficiency (50 GFLOPS/W) makes this architecture a strong candidate for all applications that do significant signal processing in embedded and mobile environments. We have exemplified the use of Epiphany in two such applications, radar and video processing. We have furthermore looked at various high-level languages based development approaches including OpenCL, Erlang, APL, and CAL. Finally, we look beyond the current generation of Epiphany and discuss what additional architectural features can be expected in future generations of Epiphany.

ACKNOWLEDGEMENT

A big thank you goes to the 4,965 Kickstarter backers who enabled the creation of the Parallella computer.

This work was supported in part by the ESCHER Project funded by the Swedish Knowledge Foundation, the HiPEC project funded by the Swedish Foundation for Strategic Research, and the ELLIIT Strategic Research Initiative funded by the Swedish Government.

The Epiphany™, eCore™, eNode™, eMesh™, eLink™, eHost™, and eLib™ are trademarks of Adapteva Inc.